\documentclass[preprintnumbers,amsmath,amssymb,nofootinbib,superscriptaddress,12pt]{revtex4}
\usepackage{amsmath,latexsym}
\begin{document}

\title{\large ON GENERATING SOME KNOWN BLACK HOLE SOLUTIONS}
\author{\large Farook Rahaman$^{1\ast}$, Mubasher Jamil$^\dag$, Ashis Ghosh$^\ast$ and Kausik Chakraborty$^\ddag$\\
\small $^{\ast}$Department of Mathematics, Jadavpur University,
Kolkata$-$700032, India\\$^\dag$Center for Advanced Mathematics and
Physics, National University of Sciences and Technology,
\\Peshawar Road, Rawalpindi, 46000, Pakistan\\$^\ddag$ Department
of Physics, Netaji Nagar College for Women, Regent Estate,
Kolkata$-$700092, India
\\$^{1}$farook[underscore]rahaman@yahoo.com\\$^\dag$mjamil@camp.edu.pk}
\begin{abstract}
\textbf{ABSTRACT:} In this paper, we have presented an algorithm to
generate various black hole solutions in general relativity and
alternative theories of gravity. The algorithm involves few
dimensional parameters that are assigned suitable values to specify
the required black hole.
\end{abstract}
 \maketitle
\newpage
\section{Introduction}

A black hole, as known since several decades, is formed as a result
of continued gravitational collapse of matter, consisting of a
singularity covered by an event horizon. It represents a region of
spacetime where a test particle is trapped in such a way that its
escape speed exceeds the speed of light. A black hole is formed when
a pressureless cloud of dust with a spherical symmetry, undergoes a
collapse which is governed by gravity only. In this scenario, the
pressure gradients are ignored and the collapse proceeds
continuously till a singularity (refers to curvature singularity only, here and onwards) is formed. Under the assumptions of
spherical symmetry, vanishing pressure gradients and asymptotic
flatness, the formation of event horizon becomes inevitable. If any
of these conditions is not satisfied, the collapse leads to a naked
singularity where the horizon either never forms or it shrinks down
to the singularity. The concept of a stationary black hole came from
Einstein's theory of general relativity. The very first solution of
Einstein field equations derived by K. Schwarzschild represented a
point source of infinite gravitational field in an empty
asymptotically flat spacetime \cite{wald,penrose}. Later on, people
constructed other black hole solutions with the inclusion of extra
parameters like charge and angular momentum \cite{hawking} (see also
\cite{stephani} for a thorough collection and history of black hole
solutions).

Since the genesis of black hole paradigm, people have been
interested in the quest of answering whether naked singularities
(singularities without horizons) could exist or not. Penrose
\cite{penrose,penrose1} put forth a `cosmic censorship conjecture'
which suggested that singularities must always be hidden inside
horizon. Although this conjecture has neither been proved nor
disproved yet has got much attention. If naked singularities
exist, then these will be observable to distant observers and
hence will be effective sources to test Einstein's theory of
relativity, in general and other theories of quantum gravity, in
particular. In a typical gravitational collapse model, the horizon
forms first and the singularity later, so that horizon
encapsulates the singularity. A naked singularity could form if
the collapse is so slow that the formation of a horizon is delayed
\cite{joshi}. Moreover, a naked singularity can also form if the
gravitational collapse is not spherically symmetrical
\cite{shapiro1}. In the context of phantom cosmology, a naked
singularity is also formed when a black hole accretes phantom
energy which possesses negative energy density. Due to its
accretion, the mass of black hole decreases eventually leading to
the appearance of a naked singularity \cite{babichev,jamil}. There
has been pioneering work done by Virbhadra and his collaborators
regarding the observational properties and detection of black
holes and naked singularities using gravitational lensing
techniques \cite{vir7,vir8,vir1,vir2,vir3,vir5,vir6}. They have
classified naked singularities into two kinds: `weakly naked'
which were hidden under single photon sphere and the `strongly
naked' which were hidden under no photon sphere. Interestingly,
the qualitative features of gravitational lensing due to a weakly
naked singularity turned out to be similar to the Schwarzschild
black hole.

In this article, we are trying to provide a prescription how to get
some known black holes existing in literature by monitoring
transverse pressure only. That means one requires the knowledge of
one of the components of energy momentum tensor, more exactly, the
transverse pressure to generate some of the known black holes,
namely, Schwarzschild, Schwarzschild de Sitter,
Reissner-Nordstr\"{o}m, Reissner-Nordstr\"{o}m de Sitter black holes
etc. Recently several authors have proposed several algorithms to
obtain spherically symmetric solutions as well as black hole
solutions \cite{kg1,kg2,kg3,kg4,kg5,kg6,kg7,kg8}.  Our approach is
simple and interesting because the algorithm involves few
dimensional parameters that are assigned suitable values to generate
the required black hole.

\section{Basic equations and the Algorithm}

Let  us consider an anisotropic matter distribution corresponding
to the line element
\begin{equation}
ds^2=  - e^{\nu(r)} dt^2+ e^{-\nu(r)} dr^2+r^2(
d\theta^2+\text{sin}^2\theta d\phi^2). \label{Eq3}
\end{equation}
The Einstein field equations for the above spherically symmetric
metric  are given by
\begin{eqnarray}
-e^{\nu}\left[\frac{\nu^\prime}{r} + \frac{1}{r^2}
\right]+\frac{1}{r^2}&=& 8\pi \rho, \\
-e^{\nu}
\left[\frac{1}{r^2}+\frac{\nu^\prime}{r}\right]+\frac{1}{r^2}&=&
8\pi
p_r ,\\
\frac{1}{2}e^{\nu} \left[(\nu^\prime)^2+ \nu^{\prime\prime} +
\frac{2\nu^\prime}{r}\right] &=&8\pi p_t.
\end{eqnarray}
Here $\rho$, $p_r$ and $p_t$ are energy density, radial and
transverse pressures, respectively. The prime denotes derivative
with respect to parameter $r$.

Eqs. (2) and (3) imply
\begin{equation} \rho = p_r.
\end{equation}
Eq. (4) yields
\begin{equation}
 2r^2 (8\pi p_t) = \frac{d}{dr} \left(r^2 e^{\nu}
 \nu^{\prime}\right).
\end{equation}
Integrating the above equation, one obtains
\begin{equation}
  r^2 (e^{\nu})
^{\prime} = \int 2r^2 (8\pi p_t)dr+D.
\end{equation}
Here $D$ is an integration constant. Integrating the above equation
once more, one gets
\begin{equation}
e^{\nu}= E - \frac{D}{r} + \int \left[\frac{1}{r^2} \int 2r^2 (8\pi
p_t)dr\right] dr,
\end{equation}
where $E$ is an integration constant. Thus one gets black hole
solutions by monitoring only one function $p_t$.

Let us assume the form of $p_t$ as
\begin{equation}
8\pi p_t = A + \frac{Q}{r^n},
\end{equation}
where $A$, $Q$ and $n$ are arbitrary constants. Using the above form
of $p_t$ in equation (8), one gets
\begin{equation} e^{\nu}
 = E - \frac{D}{r} + \frac{A}{3} r^2 + \frac{2Q}{(n-2)(n-3)
 r^{n-2}}.
\end{equation}
Eq. (2) yields
\begin{equation} 8\pi \rho= -A + \frac{(1-E)}{r^2} +
\frac{2Q}{(n-2)r^n}.
\end{equation}
Thus we obtain a general class of black hole solutions supported by
the anisotropic fluid  distribution given in (9) and (11). Here one
can note that $n \neq 2, 3$. For $n=2$ and $n=3$, one has to use
equation (8) directly to obtain metric coefficient.

\subsection{Schwarzschild Black hole}

A Schwarzschild black hole represents a point source of mass $M$ of
infinite gravitation in an empty spacetime. The spacetime is
spherically symmetric and possesses a curvature singularity at $r=0$
which is hidden inside an event horizon located at $r=2M$. This
solution is obtained from our algorithm if we put $A = Q = 0$, $D =
2M $ and $E =1$, then the above solutions imply $p_r=p_t=\rho=0$ and
$ e^{\nu} = 1 - \frac{2M}{r}$. Vanishing pressures and densities
imply that the exterior of a black hole is nothing but vacuum. This
solution is highly idealized since most black holes are rotating and
possibly surrounded by matter like stars and dust.

\subsection{Schwarzschild - de Sitter Black hole (SdS)}

A SdS spacetime represents a Schwarzschild black hole immersed in
cosmological constant (generally denoted by $\Lambda$) dominated
universe. The parameter $\Lambda$ is an ad hoc term added in the
Einstein field equations to obtain a matter distribution having
negative pressure. This matter is usually termed as the dark energy
which is homogeneous and isotropic perfect fluid. In the presence of
this fluid, the spacetime outside the black hole expands with
acceleration. If we choose $ Q = 0$, $D = 2M $ and $E =1$, then the
above solution implies $8\pi p_r=-8\pi p_t=8\pi \rho=-A$ and $
e^{\nu} = 1 - \frac{2M}{r} + \frac{A}{3} r^2 $. Here, $A$ plays the
role of cosmological constant.

\subsection{Reissner-Nordstr\"{o}m Black hole (RN)}

The RN black hole represents a spherically symmetric spacetime
containing a mass $M$ and charge $Q$. The spacetime is singular at
$r=0$ hidden under two horizons $r_{h\pm}=M\pm\sqrt{M^2-Q^2}$. If
$Q^2=M^2$, then the spacetime represents an extreme RN black hole
while if $Q^2>M^2$, it yields a naked singularity at $r=0$. If we
adopt $ A = 0$, $D = 2M $, $n =4$, $Q>0 $ and $E =1$, then the above
solutions imply $8\pi p_r=8\pi p_t= 8\pi \rho= \frac{Q}{r^4}$ and $
e^{\nu} = 1 - \frac{2M}{r} + \frac{Q}{ r^2 } $. One can note that
the above solutions correspond to braneworld black hole for $Q<0$.
From the astrophysical point of view, charged black holes are not of
much interest since these get neutralized by interacting with the
neighboring charged clouds and plasmas.

\subsection{Reissner-Nordstr\"{o}m - de Sitter Black hole (RNdS)}

The RNdS spacetime represents a RN black hole in a cosmological
constant dominated universe. The spacetime is singular at $r=0$
hidden under multiple horizons. If we take $D = 2M $, $n =4$ and $E
=1$, then the above solutions imply $8\pi p_t= A + \frac{Q}{r^4} $
and  $ 8\pi p_r = 8\pi \rho= -A + \frac{Q}{r^4}$ and $ e^{\nu} = 1 -
\frac{2M}{r} + \frac{Q}{ r^2 }+\frac{A}{3} r^2 $. For $A<0 $, case 1
and 2 indicate anti de Sitter type solutions. Thus sign of A gives
dS/AdS black holes.

\subsection{Black hole surrounded by quintessence}

Recently, Kiselev \cite{Kis} have obtained a new black solution
surrounded by quintessence. The quintessence is represented by a
homogeneous and time dependent scalar field. The equation of state
parameter $\omega_q$ of quintessence varies between $-1$ and $-1/3$.
Since $\omega_q<0$, it serves as an alternative candidate for dark
energy. If we insert $D = 2M $, $A=0$, $n =3( \omega_q +1)$, $Q=
\frac{6c \omega_q(3\omega_q +1)}{2} $ and $E =1$, then the above
solutions imply $8\pi p_t= \frac{3c \omega_q(3\omega_q +1)}{2r^{3(
\omega_q +1)}} $ and  $ 8\pi p_r = 8\pi \rho= \frac{c (3\omega_q
+2)}{r^{3( \omega_q +1)}}$ and $ e^{\nu} = 1 - \frac{2M}{r} +
\frac{c}{r^{(3 \omega_q +1) }}$.

\section{Further Observations}

Our approach is interesting in the sense that the transverse
pressure  would give similar black hole spacetimes like modified
theories of Einstein's theory give. We provide few examples:

\subsection{Example}

Beato et al \cite{boeto} have obtained a new black hole solution
using nonlinear electrodynamics as
\begin{equation} e^{\nu}
 = 1 - \frac{2m[1-\tanh(\frac{q^2}{2mr})]}{r}.
\end{equation}
Here $m$ and $q$ are respectively the mass and charge of the black
hole. Using our approach, one will see the following form of the
transverse pressure gives the very similar black hole spacetime.
\begin{equation}8\pi p_t =
\frac{1}{\cosh^2(\frac{q^2}{2mr})r^4}\left[ q^2 + \frac{q^4}{4mr}
\tanh(\frac{q^2}{2mr})\right].
\end{equation}
The other components of the energy stress tensor are given by
\begin{equation} 8\pi p_r = 8\pi \rho= \frac{1}{r^2} -
\frac{q^2}{r^4[\cosh(\frac{q^2}{2mr})]^2 } .\end{equation}

\subsection{Example}

Bardeen \cite{bardeen} have obtained  non singular black hole
solution
 as
\begin{equation}
 e^{\nu}
 = 1 - \frac{mr^2}{(r^2 +q^2)^{\frac{3}{2}}}.
\end{equation}
As above $m$ and $q$ are respectively the mass and charge of the
black hole. Using our approach, one will see the following form
of the transverse pressure provides the very similar black hole
spacetime.
\begin{equation}
8\pi p_t = \frac{m}{2(r^2
+q^2)^{\frac{7}{2}}}\left[10r^4-5q^2r^2-6q^4\right].
\end{equation}
Here, the other components of the energy stress tensor are found
to be
\begin{equation}
 8\pi p_r = 8\pi \rho= \frac{3mq^2}{2(r^2 +q^2)^{\frac{5}{2}}}.
  \end{equation}
\subsection{Example}
Dymnikova \cite{dymn} have obtained another non singular black hole
solution as
\begin{equation} e^{\nu}
 = 1 - \frac{2m[1-e^{(-\frac{r^3}{2mr_0^2})}]}{r}.
\end{equation}
Here, $m$ is mass  and $r_0^2 = \frac{3}{\Lambda} $, $\Lambda$ is
related to the positive cosmological constant. Using our
approach, one will see the following form of the transverse
pressure gives the very similar black hole spacetime.
\begin{equation}
8\pi p_t=\frac{3}{r^2}e^{-\frac{r^3}{2mr_0^2}}\left[
\frac{3r^3}{4mr_0^2} - 1\right].
\end{equation}
Here, the other components of the energy stress tensor are
obtained as
\begin{equation}
8\pi p_r = 8\pi \rho= \frac{3}{r^2}e^{-\frac{r^3}{2mr_0^2}}.
\end{equation}
\subsection{Example}
Chamblin et al \cite{cham} have obtained charged brane world  black
hole solution as
\begin{equation}
e^{\nu}
 = 1- \frac{2GM}{r}+\frac{Q^2 + \beta}{r^2} +
\frac{l^2 Q^4}{20 r^6}.
\end{equation}
Here, $M$ and $Q$ are respectively the mass and charge of the
black hole. $l$, $\beta$ are related to the bulk cosmological
constant and five dimensional mass parameter respectively. Using
our approach, one will see the following form of the transverse
pressure gives birth the very similar black hole spacetime
\begin{equation}
8\pi p_t = \frac{Q^2 + \beta}{r^4} - \frac{3 l^2 Q^4}{4r^8}.
\end{equation}
The other components of the energy stress tensor are
\begin{equation}
8\pi p_r = 8\pi \rho= \frac{Q^2 + \beta}{r^4} - \frac{ l^2
Q^4}{4r^8}.
\end{equation}

\subsection{Example }

Virbhadra et al \cite{vir} provided a conformal scalar dyon black
hole which is an exact solution of Einstein - Maxwell field
equations and is characterized by the scalar charge ($q_s$),
electric charge ($q_e$) and magnetic charge ($q_s$) as

$ e^{\nu} = (1 - \frac{Q_{CSD}}{r})^2$, where $Q_{CSD}^2 = q_s^2 +
q_e^2 + q_m^2. $

Our approach shows that  the following form of the transverse
pressure gives the very similar black hole spacetime.

\begin{equation}
(8\pi p_t) = \frac{Q_{CSD}^2}{r^4}.
\end{equation}
Here, the remaining components of the energy stress tensor are
\begin{equation}
8\pi p_r = 8\pi \rho = \frac{Q_{CSD}^2}{r^4}.
\end{equation}

\section{Final remarks}

In this paper, we have presented a scheme of generating some known
spherically symmetric black hole solutions satisfying Eq. (1).
Other interesting solutions like Kerr, Kerr-Neumann and
Janis-Newman-Winicour solutions \cite{vir4} are not obtained from
our scheme. Though we do not provide any new black hole solutions,
despite we give a clue how one can get some black hole spacetimes.
We would also mention that the solutions discussed in this paper
may not necessarily black holes and could be naked singularities.
The differentiation between these solutions can decisively be
obtained through more rigorous analysis. Our approach may be
applied to generate various black hole solutions in general
relativity and alternative theories of gravity \cite{nico}.

\subsection*{Acknowledgments} FR is thankful to
Jadavpur University and UGC, Government of India for providing
financial support. FR is also grateful to IMSc, for providing
research facilities. We would like to thank the anonymous referee
for giving critical comments on this work.

\end{document}